\newcommand{\ig}[1]{

In this article r = length  r_min = 1/Lambda QCD.
and R r_min is dimensionless!

In my kitp notes ``r'' = r/R^2 has unit of mass!

}
\documentclass[12pt]{article}
\usepackage{epsfig}

\newcommand\be{\begin{equation}}
\newcommand\ee{\end{equation}}
\newcommand\bea{\begin{eqnarray}}
\newcommand\eea{\end{eqnarray}}
\newcommand\nn{ \nonumber\\}

\newcommand{\text}[1]{\qquad \mbox{#1} \qquad}

\newenvironment{myQuote}[2]{\begin{list}{}{\leftmargin#1\rightmargin#2}\item{}}{\end{list}}

\newcommand{\tbox}[1]{\qquad \mbox{#1} \qquad}

\newcommand{\<}{\langle}
\renewcommand{\>}{\rangle}

\title{String/Gauge Duality:\\
(re)discovering the QCD String in AdS Space
\thanks{Based on three lectures at the 43rd Cracow School of Theoretical
  Physics, Zakopane, Polan, June 2003.}
}

\author{Richard C. Brower
Physics Department\\
Boston University\\
590 Commonwealth Ave, MA 02215}

\begin{document}

\maketitle

\begin{abstract}
These lectures trace the origin of string theory as a
theory of hadronic interactions (predating QCD itself) to the present
ideas on how the QCD string may arise in Superstring theory in a
suitably deformed background metric. The contributions of 'tHooft's large
$N_c$ limit, Maldacena's String/Gauge duality conjecture and lattice
spectral data are emphasized to motivate and hopefully guide further
efforts to define a fundamental QCD string.
\end{abstract}
 
\section*{Preface: Not by accident}

String theory, contrary to conventional lore, was discovered not by accident
but by a systematic program to build a relativistic quantum theory of the
hadronic interactions without resorting to the use of local fields.  The
approach, referred to as ``S matrix theory'', sought to impose a minimal set
of consistency conditions directly on the S matrix~\cite{note}.  At the time,
it appeared absurd to consider the known light hadrons (pions, nucleons,
etc.) as ``elementary'' fields, particularly with the realization that they
were just the first member of a Regge family of increasingly higher masses and
spins ($J \simeq \alpha' m^2_J + \alpha_0$). In the language of low energy
effective field theory, the difficulties in formulating a quantum field theory
of hadrons and gravity were analogous.  The effective low energy theory of
hadrons (e.g. the pions) is the chiral Lagrangian,
\be
S[U] = \int d^4x \{\frac{F_\pi^2}{4} \mbox{Tr}
[\partial_\mu \Sigma^\dagger \partial_\mu \Sigma] -
\frac{\langle \bar \Psi \Psi \rangle}{2 N_f} \mbox{Tr}
[{\cal M} \Sigma^\dagger + {\cal M}^\dagger \Sigma] + \cdots \} \; ,
\ee
and for gravity the Einstein-Hilbert Lagrangian,
\be
S[g]  =   \frac{M^2_P}{16 \pi^2} \int d^4x \{ \sqrt{-g} \;  ( R + \Lambda) + \cdots  \}
\; .
\ee
Both are beautiful geometric quantum theories, but they are
non-renormalizable with dimensionful coupling constants inversely
proportional to mass ($1/F_\pi$ and $1/M_P$). In each order of the loop
expansion, one must cancel UV divergences with new higher dimensional counter
terms. With the advent of QCD the analogy appeared to be lost. But it is the
goal of these lectures to argue that this is not the case.

Due to dimensional transmutation QCD (with massless quarks) has a single
fundamental mass scale, $\Lambda_{qcd}$, but no coupling constant. Consequent
the only available ``perturbative'' expansion for QCD (in the infrared) is
the 'tHooft expansion for small $1/N_c$ at fixed $\Lambda_{qcd}$.  This
expansion leads to a distinctly string like hadronic phenomenology. However
the central question of these lectures is {\bf not } the obvious existences
of a phenomenological QCD string but the more basic question:
\begin{myQuote}{0.25in}{0.25in}
\sc
Is the Yang Mills theory for QCD exactly equivalent (i.e. dual) to
a fundamental String Theory?
\end{myQuote}
This question goes beyond the existence of a confining QCD vacuum with
stringy electric flux tubes to the question of a mathematically precise
identity between QCD and string theory in the same sense that the Sine Gordon
and Massive Thirring quantum theories are equivalent. In the latter example, not
only does duality exchange strong and weak coupling expansions, but after all
non-perturbative effects are included the Sine Gordon and Massive
Thirring theories have identical S matrix.

For many years a similar identity between QCD and some form of string theory
has been sought. At long last, recent progress in superstring theory gives
this endeavor a more concrete form. Based on Maldacena's AdS/CFT
conjecture~\cite{Maldacena:1997re}, backed up by almost 5 years of
consistency checks, the existence of an exact Gauge/String duality between
some (super) Yang Mills theories and superstrings in a non-trivial
(asymptotically AdS) background is now generally
accepted\cite{Aharony:1999ti}. We now have concrete mathematical support for
a generic mechanism for string/gauge duality linked to the so called
``holographic principle'' for any theory including quantum gravity.
Naturally this has revived the search for a QCD string and brought many
features into much clearer focus.  These lecture will briefly review the
history and recent progress in this ancient quest for the QCD string.
Moreover it should be added that success in constructing a hadronic string
would not only be of interest in gaining a deeper understanding of QCD but,
if successful, a major step in understanding what constitutes string theory
itself.
 
\section{Lecture One: Ancient Lore}

\subsection{Empirical Basis}

The discovery of string theory in the late 1960's followed from a detail
study of the phenomenology of hadronic scattering, specifically finite energy
sum rules constrained by Regge theory at high energies. For example the Regge
limit for pion elastic scattering amplitude ($\pi^+ \pi^- \rightarrow \pi^+
\pi^-$) was traditionally parameterized as
\be
A_{\pi^+ \pi^- \rightarrow \pi^+ \pi^-} (s,t) 
\simeq g_o^2 \; \Gamma[1- \alpha_\rho(t)] \; (- \alpha' s)^{\alpha_\rho(t)} \; ,
\ee
in Mandelstam variable $s = - (p_1 + p_2)^2$ and $t = - (p_1 + p_3)^2$.
The Gamma function prefactor gives cross channel poles for rho exchange at
J=1 and higher spins for $J > 1$. Since the ratio for the rho width to mass is a small
parameter ($\Gamma_\rho/m_\rho \simeq 0.1$), one sought a new perturbative
expansions starting with a zero width approximation. This was traditionally
enforced for all resonant states by using an exactly linear rho trajectory,
$\alpha(t) = \alpha' t + \alpha_0$, so that ``resonance'' poles at integer
$J = \alpha(m^2)$ had real masses~\cite{BrowerHarte}.  In 1968
Veneziano~\cite{veneziano} realized that exact $s,t$ crossing symmetry could
be imposed by assuming an amplitude of the form,
\be
A_{\pi^+ \pi^- \rightarrow \pi^+ \pi^-}(s,t) 
= g_o^2 \; \frac{ \Gamma[1- \alpha_\rho(t)] \Gamma[1-\alpha_\rho(s)]}{\Gamma[1- \alpha_\rho(s) - \alpha_\rho(t)] } \; ,
\label{equ:veneziano}
\ee
the so called {\bf dual} resonance model. Here ``dual'' referred to
Dolan-Horn-Schmid duality~\cite{Dolen:1967jr} which states that the sum over
s-channel resonances poles interpolates the power behavior of the leading
Regge pole exchange,
\be
 \sum_r \frac{g^2_r(t)}{s  - (M_r - i \Gamma_r)^2} \simeq \beta(t) (-\alpha'
 s)^{\alpha(t)} \; .
\ee
This property is easily derived for the dual pion scattering amplitude
(\ref{equ:veneziano}).  The Regge limit  follows from the Sterling's
approximation as $s \rightarrow \infty$ and the resonance expansion 
from the integral representation for the Beta function,
\be
A_{\pi^+ \pi^- \rightarrow \pi^+ \pi^-}(s,t)  =  - g_o^2 \alpha_\rho(t) \int^1_0 dx \; x^{- \alpha_\rho( s)}  (1-x)^{-1- \alpha_\rho( t)} \; .
\ee
Expanding at small x we get,
\bea
A_{\pi^+ \pi^- \rightarrow \pi^+ \pi^-}(s,t) &=& -g_o^2  \sum^\infty_{J=1} \frac{(\alpha_\rho(t))(\alpha_\rho(t) + 1) 
\cdots (\alpha_\rho(t) + J-1)}{(J-1)!} \int^1_0 dx \; x^{-1- \alpha( s) + J } \nn
\Rightarrow   \sum^\infty_{J=1} \frac{g_o^2 A_J(\alpha' t)}{\alpha_\rho(s) - J} 
&\simeq& g_o^2 \Gamma(-1-\alpha_\rho(t)) (- \alpha' s)^{\alpha_\rho(t)} \; ,
\eea
where  $A_J$  is  a polynomial  of  order  $J$.  In fact  the  initial
enthusiasm  for this  model included  a striking feature  of chiral
symmetry.  In the soft pion limit $p_1
\rightarrow 0$, the Adler zero,
\be
A_{\pi^+ \pi^- \rightarrow \pi^+ \pi^-}(s,t) = (1- \alpha_\rho(s) - \alpha_\rho(t)) \frac{ \Gamma[1- \alpha_\rho(t)] \Gamma[1-\alpha_\rho(s)]}{\Gamma[2- \alpha_\rho(s) - \alpha_\rho(t)] } \sim \alpha'(s + t) \rightarrow 0 \; ,
\ee
is imposed if we take the phenomenologically reasonable values for the rho
trajectory intercept, $\alpha_\rho(0) = 0.5$.  Further work led to the
N-point generalization in Neveu and Schwarz's seminal
paper~\cite{Neveu:1971rx} entitled ``Factorizable Dual Model of Pions''.  So
Veneziano's amplitude turns out to be the 4-point function of the NS
superstring ---ignoring the conformal constraint on the Regge intercept
($\alpha(0) = 1$) and the dimension of space time ($D=10$) which was not
understood at the time.

As we will explain this initial enthusiasm was premature.

\subsection{Covariant String Formulation}

It is surprisingly easy to generalize the 4-point Beta function to get the 
N-point dual resonance amplitudes and the covariant quantization of
the Bosonic string. The argument goes as follows. Consider the 4-point
function for tachyon scattering~\cite{Polchinski:1998rq}  in a symmetric form,
\be
\int^1_0 x^{-1 - \alpha(s)} (1-x)^{-1 - \alpha(t)} dx  =   \int^{x_3}_{x_1} \frac{dx_2}{(x_4 - x_3)(x_4-x_1)(x_3 -x_1)} \prod_{1\le i <j \le 4} (x_j - x_i) ^{2 \alpha' p_j p_i}  \; ,
\ee
where $\alpha(s) = \alpha's + 1 = 2 \alpha' p_1 p_2$ for $\alpha' p^2_i =
-1$. The three dummy variables maybe fixed at $x_1 = 0, x_3 = 1, x_4 =
\infty$.  This does not spoil cyclic symmetry, since the integrand is
invariant under M\"obius transformations: $x_i \rightarrow (ax_i + b)/(cx_i +
d)$. 

Now there is an obvious guess to generalize the 4-point amplitude to
N-point open string tachyon amplitude,
\be
A_N(p_1,\cdots,p_N) =   g_o^{N-2}\int\frac{dx_2 dx_3 \cdots dx_{N-2}}{(x_N -
  x_{N-1})(x_N-x_1)(x_{N-1} -x_1)} \prod_{1\le i <j \le N} (x_j - x_i) ^{2
  \alpha' p_j p_i}  \; .
\ee
The integration region is restricted to be $ x_1 \le x_2 \le x_3 \le \cdots
\le x_N$. Modern string theory lectures or textbooks usually require hundreds
of pages of derivation to write down this amplitude, if they bother to do it
at all. (This is not to imply that you should not learn the formal approach to
string path integral quantization but the discovery of string theory was in
large part due to the simplicity of the final answer for the tree amplitude.
Pedagogically it may even help to understand the answer in advance of its
derivation.)

One can also follow the pioneers of the field and write down the Old
Covariant Quantized string, working ``backward'' from the N-point function.
One needs to factorize the N-point function, i.e introduce a complete set of
states. Short circuiting the full derivation, this amounts to a free (string)
field expansion,
\be
X^\mu(\sigma,\tau) =  \hat q^\mu -  2 i  \alpha' \hat p^\mu \tau + \sum^\infty_{n=1}
\sqrt{\frac{2 \alpha'}{n}} 
(a^\mu_n \exp[- n \tau] + a^{\mu \dagger}_n \exp[n \tau] )cos(n\sigma) \;
\ee
into normal mode oscillators,
\be
[ \hat q^\mu,\hat p^\nu] = i  \eta^{\mu\nu} \tbox{and} 
[a^\mu_n, a^{\nu \dagger}_m] = \eta^{\mu \nu} \delta_{n,m} \; ,
\ee
acting on the ground state tachyon at momentum $p$,
\be
\hat p^\mu  | 0,p\> = p^\mu |0,p\>  \tbox{and}  a^\mu_n | 0,p\> = 0 \; .
\ee
Then a short algebraic exercise will convince you that the integrand for the 
N-point function does factorize as,
\be
\<0,p_1|V(x_2,p_2) V(x_3,p_3) \cdots V(x_{N-1},p_{N-1}) |0,p_N\> = \prod_{1 \le
  i < j \le N} (x_j - x_i)^{ 2 \alpha' p_j p_i}
\ee
with
\bea 
V(x,p) &=&  :\exp[i p X(0,\tau)]: \\  &=& 
\; \exp[i p \hat q + ip \sum_n \sqrt{\frac{2 \alpha'}{n}}   a^\dagger_n x^{n}
] \; \exp[2 \alpha' p\hat p \log(x) + i p
\sum_n \sqrt{\frac{2 \alpha'}{n}}   a_n
  x^{-n} ] \nonumber
\eea
and $x \equiv \exp[  \tau]$. To calculate the matrix element (i.e amplitude)
one merely normal orders the operators giving factors,
\be
\exp[ - 2 \alpha' \sum_n \frac{p_i p_j}{n} (\frac{x_i}{x_j})^n
] = \exp[ \alpha'p_i p_j\log(1 - x_i/x_j)]
= (1 - \frac{x_i}{x_j} )^{p_ip_j} \; ,
\ee
for each pair of vertex insertions.  The stringy interpretation follows from
identification of world sheet surface co-ordinates $(\sigma,\tau)$. The above
expansion for the space-time position, $X^\mu(\sigma,\tau)$, is a solution to
the 2-d conformal equations of motion for a free string,
\be
  \partial_\tau^2 X^\mu +\partial_\sigma^2X^\mu  = 0  \; ,
\label{eq:eom}
\ee
in Euclidean world sheet metric. Writing down the general normal mode
expansion,
\be
X^\mu(z,\bar z) = \hat q^\mu -  i \alpha' \hat p^\mu log(z\bar z) +
\sum^{\infty}_{n =-\infty,n \ne 0}
 \sqrt{\frac{2 \alpha'}{|n|}} ( a^{\mu} _n z^{-n} + b^{\dagger \mu}_n {\bar z}^{-n} )
 \; ,
\ee
with $z = \exp[\tau + i \sigma]$, we see that the particular solution required
for the open string amplitude above satisfies Neumann boundary conditions at
the ends, $\sigma = 0, \pi$. The vertex function representing tachyon
emission are inserted on one side at the $\sigma =0$ boundary.  (Closed
string have periodic boundary conditions in $\sigma$. Super strings add
worldsheet 2-d fermion fields. That's it. Sum over all worldsheet
Riemann surfaces, with great care, and you have (perturbative) superstring
theory.)

Nambu and Gotto took the stringy interpretation of the dual model one step
further by noticing that the equation of motion (\ref{eq:eom}) is a gauge
fixed form for a general co-ordinate invariant world sheet (Nambu-Gotto)
action, \be S_{NG} = - \frac{1}{2\pi \alpha'} \int d^2\xi \sqrt{ - det (h)}
\;, \tbox{where} h_{\alpha \beta} = \partial_\alpha X^\mu \partial_\beta
X_\mu \; , \ee
with surface tension $T_0 = 1/(2 \pi \alpha')$. At the classical level
this is also equivalent to the Polyakov form,
\be
S_{P} =  - \frac{1}{2\pi \alpha'} \int d^2\xi
\sqrt{ - det(\gamma)} [\gamma^{\alpha \beta} \partial_\alpha X^\mu
\partial_\beta X_\mu ]  \; ,
\ee
with an auxiliary ``Lagrange multiplier'' 2-d metric, $\gamma_{ij}$.
However the Polyakov form is easier to gauge fix and quantize using BRST
technology~\cite{Polchinski:1998rq}. To get a feeling for the dynamics of the
open string, it is interesting to write down a few classical
solutions~\cite{Zwiebach:2004tj}.

\subsection{Two Open String Solutions}

One can write down the Euler Lagrange equations for the Nambu-Gotto string
action and use diffeomorphism invariance, $\tau' =f(\sigma,\tau), \sigma' =
g(\sigma,\tau)$, to choose a gauge. The static ($t = X^0 = i\tau$) orthogonal
($h_{12} = h_{11} + h_{22} = 0$) gauge  is a useful choice. This gives a
linearize equation of motion 
\be \partial_t^2 X^\mu - \partial_\sigma^2X^\mu = 0 \; , 
\ee
with constraints,
\be
\partial_\sigma X^\mu \partial_t X_\mu = 0
\qquad
 \partial_t X^k \partial_t X_k +  \partial_\sigma X^k \partial_\sigma X_k = 1
 \; .
\ee

\underline{\bf Solution \# 1:} The string stretch along the 3rd axis with
(fixed) Dirichlet boundary conditions, $\sigma \in [0,L]$: All spatial
components $X^k = 0$ except,
\be
X^3 = \sigma \; ,
\ee
with energy $E_0 = T_0 L$ exhibiting linear confinement.  For future
reference the exact quantum solution has energy,
\be 
E_n = T_0 L \sqrt{ 1 - \frac{\pi(D-2)}{12 T_0 L^2} + \frac{2 \pi
    a^\dagger_n a_n}{T_0 L^2}} 
\ee
for D space-time dimensions.

\underline{\bf Solution \# 2:}  The free string rotating in the $(X_1, X_2)$
plane with Neumann boundary conditions, $\sigma \in [0, \pi L/2]$: All spacial
components  $X^k = 0$ except, 
\be
X^1 + i X^2 = (L/2) \cos(2 \sigma/L) \exp[i2 t/L] \; ,
\ee
with energy $E = \pi L T_0/2$ and total angular moment (spin) $J =
\alpha' E^2$. This last result ($J \sim E^2$), which is the key
requirement for QCD Regge phenomenology, is a rather non-trivial
property of a relativistic massless string. The end points always
travel at the speed of light, so as the energy increases the string
gets longer BUT the angular velocity decreases: $\omega = 2/L = 1/(\pi
T_0 E)$. Nonetheless the angular momentum increases quadratically
because at constant tension the total stored energy grows linearly in
L and the moment of inertia grows as a cubic, $L^3$. This is in stark
contrast with a rigid non-relativist bar where $J \sim
E^{1/2}$. Clearly the linear Regge trajectory support the general
picture of a massless ``flux'' tube with energy coming entirely from
its tension.  Again for future reference the exact quantum state for
this leading trajectory is
\be
(a^1_{(1)} + ia^2_{(1)})^J |0,p\rangle \; . 
\ee

\subsection{Failure of the Old QCD String}

We should now take a break from this discourse and learn all of rules of
superstring perturbations theory~\cite{Polchinski:1998rq2}. With the help of
anomaly cancellation, we would discover 5 consistent perturbation expansions
--- free of tachyons and negative norm (i.e. ghost) states. The resulting
phenomenology for perturbative superstrings (in flat space-time) has 4
disasters from the view point of a QCD string:
\begin{enumerate}
\item Zero mass states (i.e $1^-$ gauge/ $2^{++}$ graviton)
\item Supersymmetry
\item Extra dimension: $4 + 6 = 10$
\item No Hard Scattering Processes
\end{enumerate}

One can easily imagine that the first 3 difficulties could be remedied by
``forcing'' some form of compactification of the extra 6 dimensions, breaking all
unwanted symmetries. Indeed in view of the fact that superstrings include
gravity, it is even natural to suppose that solutions should include
non-trivial space-time geometries. However the 4th problem  (no hard
scattering) reveals a fundamental mismatch between soft strings and hard
partonic QCD.  All in all an abject failure for QCD strings -- albeit a very
interesting framework for a theory of quantum gravity interacting with
matter. A theory of Everything perhaps. There are two possible consequence,
either the fundamental QCD string has nothing to do with a fundamental
superstring or there are dramatic new effects when non-trivial background
metrics are considered.

\section{Lecture Two: Gauge/String Duality }

In a sense the modern era of the QCD string begins almost immediately after
the discovery of QCD itself with 'tHooft analysis~\cite{thooft} of the large
$N_c$ limit in 1974. The problem he faced was to understand how the picture
of valence quarks attached to the strings of the dual resonance model might
arise in QCD.  Even assuming some non-perturbative mechanism for electric
confinement, one must find a small parameter to explain the zero resonance
width approximation.

\subsection{Large $N_c$ Topology}

Note that full quantum theory for QCD has in fact no coupling constant
because by dimensional transmutation (or breaking of conformal symmetry at
zero mass for the quarks) this coupling is replaced by a fundamental mass
scale, $\Lambda_{qcd}$.  Thus SU(3) Yang Mills theory has in fact no free
dimensionless parameters relative to the intrinsics QCD scale
$\Lambda_{qcd}$, except for the masses of the quarks $m_q/\Lambda_{qcd}$ and
the $\theta$ parameter.  The so called weak coupling expansion for QCD, is a
shorthand for the loop expansion in $\hbar$ which is of course of great use
in the UV for large ``energies'', $E/\Lambda_{qcd}$, due to ``asymptotic
freedom''.

Consequently, 'tHooft asked whether the inverse of the rank of the group for
$SU(N_c)$ Yang-Mills theory could be used as a formal expansion parameter.
Term by term in the loop expansion in $\hbar$, he suggested expanding in
$1/N_c$ holding fixed the 'tHooft coupling $g^2_{YM} N_c$.  Resumming each
contribution to $(1/N_c)^n$, the result is the famous topological
restructuring of the loop expansion as sum over Riemann surfaces.  The
derivation proceeds as follows. Starting  from the action,
\be
  S= \frac{1}{g^2_{YM}}Tr[(\partial_\mu A_\nu - \partial_\nu A_\mu 
+ i  [A_\mu, A_\nu])^2] +  \frac{1}{g^2_{YM}}\bar \Psi(\gamma_\mu
  \partial_\mu - i  A_\mu) \Psi \; ,
\ee
we write down Feynman expansion tracing the color and flavor flow in the
``double line'' diagramatics and count factors of $1/N_c$:
\bea
\mbox{Gluon Loops} &:&  \delta^r_r = N_c \; \Rightarrow O(N_c^F) \nn  
\mbox{Gluon \& Quark Prop}  &:& g^2_{YM} = g^2_{YM} N_c \times \frac{1}{N_c} \; \Rightarrow \; O(N_c^{-E}) \; ,  \nn
\mbox{Vertices}  &:& \frac{1}{g^2_{YM}} =  \frac{1}{g^2_{YM}N_c} \times N_c
\Rightarrow  \; O(N_c^V)  \; .
\eea
Using Euler's theorem the factors of $N_c$ for color loops (faces F),
gluon/quark propagators (edges E), interactions (vertices V) is rewritten and
quark flavor loops (boundaries B),
\be
N_c^{F - E + V - B} = N_c^{\chi}  = N_c^{2 - 2 H - B} \; ,
\ee
depending only on the topology of the graph as function of the number of
glueballs propagators (i.e. handles H) and the quark loops (
i.e.boundaries B). This is precisely the topological expansion of string
theory in terms of the genus of the world sheet. 

Perhaps more significant this topology can also be shown to hold on the
lattice in the strong coupling confining phase.  On the lattice the strong coupling expansion
is actually a sum over surfaces of electric flux so in spite of the extreme
breaking of Lorenz invariance due the lattice, the physical mechanism for
confinement is clearly string-like flux tubes.  The derivation is 
analogous to weak coupling. For illustration consider the Wilson form of the
pure gauge action,
$$
S= \frac{1}{g^2_{YM}} \sum_P Tr[2 - U_P - U^\dagger_P] \quad , \quad      
U_P = U_\mu(x) U_\nu(x+\mu) U^\dagger_\mu(x+\nu) U^\dagger_\nu(x) \quad , \quad      
U_\mu = \exp[i a A_\mu] \; .
$$
as a sum over plaquettes. In strong coupling the action is expanded in a
power series and each link variable ($U_\mu(x)$) is integrated over its Haar
measure. To get a non-zero result every link in the expansion must be paired
with (at least) one anti-link ($U \rightarrow U^\dagger $). This leads
immediately to the rule:
\bea
\mbox{Plaquettes} &:& \frac{1}{g^2_{YM}} =  \frac{1}{g^2_{YM}N_c} \times
N_c\; \Rightarrow O(N_c^F) \; , \nn  
\mbox{Links}  &:& \int dU \; U^{l_1}_{r_1} U^{\dagger r_2}_{l_2}  
= \frac{1}{N_c} \delta^{l_1}_{l_2}  \delta^{r_2}_{r_1} \; \Rightarrow \; O(N_c^{-E}) \; ,  \nn
\mbox{Sites}  &:& \delta^r_r = N_c \; \Rightarrow  \; O(N_c^V)  \; ,
\eea
Treating quark loops boundaries as before, Euler's theorem yields {\bf
exactly} the same topological result as in weak coupling (ignoring
self-intersections of surfaces). However it should be realized that the
meaning is quite different.  The vertices give the index sums, the faces are
now field strengths and edges are not propagators. Apparently the topology of
large $N_c$ Yang Mills is a robust feature in need of a deeper explanation.

In a real sense the large $N_c$ limit if it exists can be considered as one
possible {\bf definition} of the QCD string perturbative expansion order by
order in the string coupling $g_s \sim 1/N_c$.  But to go beyond this
theoretical assertion by explicitly take the large $N_c$ limit to give a
mathematical tractable formulation of the perturbative QCD string (even for
the leading term at $N_c = \infty$) has proven frustrating, except for two
dimensional QCD.  Also it is interesting to note that there is more than one
large $N_c$ limit~\cite{Armoni:2003fb}. One can choose to treat quark field
as an anti-symmetric tensor, $\Psi^{ij} = \epsilon^{ijk} \psi_k$ in color. If
one now takes the large $N_c$ limit of 1 flavor QCD with this tensor
representation for quark fields, the fermion loop is no longer subdominant.
Now the leading term can be shown to be precisely the same as the large
$N_c$ limit of ${\cal N} = 1$ SUSY Yang Mills theory!  Should we be alarmed
at this in view of the glib statement that the large $N_c$ limit defines
string perturbation theory. I think not.  In fact the full non-perturbative
QCD string theory might well have more than one weak coupling string
expansion, analogous to the now conventional view of superstrings in 10-d.

\subsection{AdS/CFT correspondence for Superstrings}

String theory has undergone a tremendous transformation in the last 35 years.
In the ``First String Revolution'' perturbative string vacua were restricted
to five alternatives (IIA, IIB, I, H0, HE) by the requirement to  cancel
tachyons, ghosts and anomalies. This appeared to restrict dramatically
the space of possible string theory. In the ``Second String Revolution'',
non-perturbative dualities even related these 5 cases (and M theory) into a
single connected manifold.  However, that is not the end of the story.
Solitonic objects called D-brane have given rise to a tremendous explosion of
possible vacua  so in the infrared the physics of strings in non-trivial
backgrounds are seen to mimic a plethora of effective fields theories.

In 1998 Maldacena~\cite{Maldacena:1997re} realized that at least under certain
circumstances string theories had to be dual (i.e. equivalent) to Yang Mills
theory. While this is technically still a conjecture, consistency relations
are now so extensive that the existence of exact String/Gauge dualities in
many special circumstances is hard to doubt.

\begin{figure}
\begin{center}
\includegraphics[width=3.0in,height=6.0in,angle=-90]{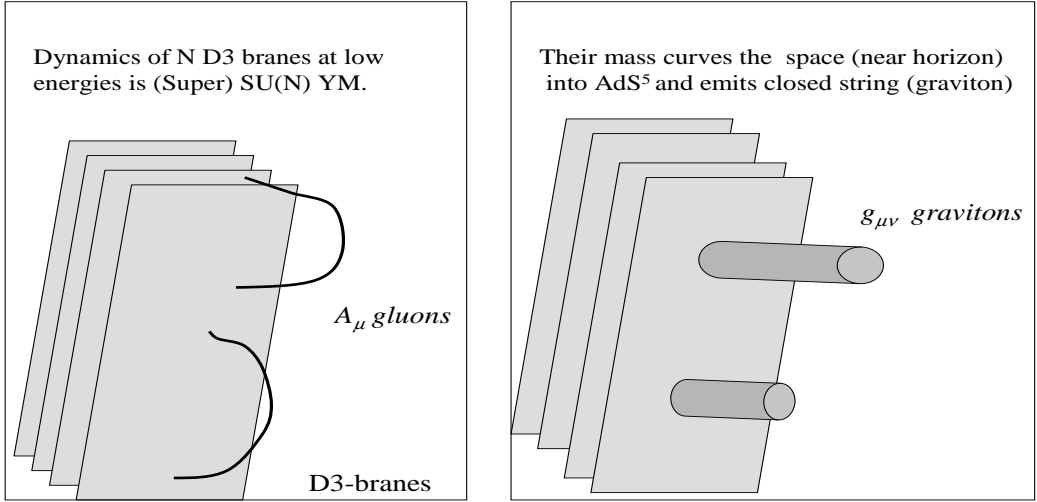}
\end{center}
\caption{ Open/closed string duality for $N_c$ D3 branes leading to 
the conjecture duality of IIB strings in $AdS^5 \times S^5$ and
$\cal N$ = 4 Super SU($N_c$) Yang Mills theory.}
\label{fig:DW}
\end{figure}

The first example was IIB superstrings (or in the low energy limit IIB
supergravity) propagating in an $AdS^5 \times S^5$ 10-d manifold,
\be
ds^2 = \frac{r^2}{R^2}\sum^3_{\mu=0} \eta_{\mu\nu} dx^\mu  dx^\nu +  \frac{R^2}{r^2}\; (dr^2   + r^2 d^2\Omega_5)
\ee
which is dual to 4-d $\cal N$ = 4 Super SU($N_c$) Yang Mills theory.  The
$(x_\mu, r)$ co-ordinates form the $AdS^5$ manifold with negative curvature
with radius $R$ fibered by a $S^5$ sphere of (positive) radius R
and metric $d^2\Omega_5$.  The motivation for this duality is
based on the background metric for a set of $N_c$ parallel massive D3 sources
(see Fig.~\ref{fig:DW}).  Evidence had accumulated that there are two
equivalent ways to model the dynamics of D3 branes. First by considering
short open strings attached to the branes which at low energies is SUSY Yang
Mills (SYM) theory  and second by the near horizon fluctuations of closed IIB
superstrings or at low energy IIB supergravity.  The leap of faith was to
conjecture that in the near horizon limit these are exactly equivalent. In
a sense this is the old open/closed string duality in a different context.

In this dual correspondence, the string (or gravity) correlation functions as
you approach the boundary of $AdS^5$ ($r \rightarrow \infty$) are equivalent
to the gauge invariant correlators in SYM theory. The discrete
``Kaluza-Klein'' modes in $S^5$ give the multiplets under SYM R symmetry
$SU(5)$. By combining the subtle new idea of holography in r and the more
mundane Kaluza-Klein mechanism on $S^5$, we see how a 10-d string can be dual
to a 4-d field theory.  There is no loss of degrees of freedom.  The 'tHooft
gauge coupling is $g^2_{YM}N_c =R^4/\alpha'^2$ where the intrinsic string
length scale is $\sqrt{\alpha'} = l_s$.  Consequently 'tHooft's strong
coupling gauge theory is dual to weak coupling gravity ($l_s \sim
l_{Planck}$) and the $1/N_c$ expansion parameter is identified
with the closed string coupling constant $g_s = g^2_{YM} \sim 1/N_c$ as one
would expect from the large $N_c$ topological expansion.

Although the Maldacena string/gauge duality is believed to hold for general
coupling and general $N_c$, it is difficult to quantize even free strings
($N_c = \infty$) in this background which includes a non-zero Ramond-Ramond
flux. In the strong coupling limit ($g_s \sim g^2_{YM} \rightarrow \infty$),
the string tension diverges leaving only the center of mass motion of closed
strings, which is equivalent to IIB gravity in the tree approximation.
The weak coupling limit of classical gravity is easily solved.
(Other special cases, such as the pp-wave limit, are tractable as well.)

\subsection{Confinement}

One may view the correspondence in holographic terms. The Yang Mills UV
(short distance) degrees of freedom are dual to excitations near to the AdS
boundary at $r \rightarrow \infty$, while the IR (long distance physics) is
represented by modes at small $r \rightarrow 0$. This mapping is referred to
as IR/UV correspondence. A graphic illustration of this IR/UV correspondence
is afforded by the scale breaking instanton solution to Yang-Mills located
at $x^\mu$ with size $\rho$. This corresponds exactly to 0-brane located at
five dimensional co-ordinate $(x^\mu, r = 1/\rho)$ in the $AdS^5$ manifold.

Ironically this first example of Yang-Mills/String duality does not confine
because the quantum field theory is exactly conformal.  Wilson loops have
pure Coulomb (rather than area law) behavior. When a large Wilson test loop
is introduced on the boundary of AdS, the red shift factor $r^2/R^2$ of the
metric allows the minimal surface area spanning the loop to remain finite by
moving into the interior nearer and nearer to $r = 0$.

To look for models closer to QCD we must break conformal and
supersymmetries. These models typically modify the metric in the IR
cutting it off at a finite value $r = r_{min}$. Two simple examples
were suggested by Witten by introducing a Euclidean AdS black hole
background with a compact dimension (called $\tau$) whose radius set by
the Hawking temperature:
\begin{itemize}
\item $AdS^5 \times S^5$ Black Hole for 10-d IIB string theory 
\item $AdS^7 \times S^4$ Black Hole for 10-d IIA string theory 
\end{itemize}
Both metrics have the general form,
\be
ds^2 = \frac{r^2}{R^2} \eta_{\mu\nu} dx^\mu  dx^\nu 
+  \frac{R^2}{r^2[1 - (r_{min}/r)^d]}\; dr^2  
+ \frac{r^2}{R^2}   [1 - (r_{min}/r)^d]\; d\tau^2 + 
  ds^2_X \; .
\ee
The horizon of the black hole introduces a scale breaking cut-off, which we
can identify roughly with $\Lambda_{qcd} = 1/r_{min}$ or as we will see
subsequently the  scale of the glueball mass in strong coupling.

\bigskip
\setlength{\unitlength}{1.00 mm}
\begin{picture}(100,65)
\linethickness{.65mm}
\put(10,35){\vector(1,0){80}}
\put(20,30){\line(0,1){10}}
\put(75,30){\line(0,1){10}}
\put(22,40){$r = r_{min}$ (IR)}
\put(85,40){$r = \infty$ (UV)}
\put(10,30){$0$}
\put(60,20){   $* \longleftarrow \longleftarrow \longleftarrow $   point defect in AdS at  $(x,r  = 1/\rho)$ }
\put(80,15){ $\Leftrightarrow$ Instanton at $x$ radius =$\rho$ }
\qbezier(20,35)(20,65)(75,65)
\qbezier(20,35)(20,5)(75,5)
\qbezier(75,65)(70,35)(75,5)
\put(80,60){$X$}
\qbezier(75,65)(80,35)(75,5)
\end{picture}

In these black hole metrics, the minimal area surface spanning a Wilson loop
of increasing size eventually must approaches $r=r_{min}$. At this point the
area of the surface no longer has a red shift factor and it grows
proportional the physical area of the Wilson loop itself.  For example in the
$AdS^5$ black hole the proper areas grows proportional to $r^2_{min} R^2$
giving a QCD tension $T_{qcd} = 1/2 \pi \alpha'_{qcd}$   or Regge slope,
\be
\alpha'_{qcd} = \alpha'/ R^2 r^2_{min} = \Lambda^2_{qcd}/\sqrt{2 g^2_{YM} N_c} \label{eq:alpha5}
\ee

\subsection{Hard Scattering at Wide Angles}

We know that QCD, even in leading order of large $N_c$, exhibit asymptotic
freedom and hard parton scattering properties.  Consequently for the QCD
string, one of the most baffling features in flat space is the complete
absence of hard scattering. One the other hand for  the application of string
theory to quantum gravity, this softening of the short distance physics is a
virtue, which is responsible for a finite weak coupling limit. Here we explain
a surprisingly simple mechanism to reconcile this apparent conflict for
strings duals to Yang-Mills theory.

Let us begin with a description of the fundamental ``Rutherford experiment''
for hadrons -- scattering two hadrons at wide angles. It is known that QCD exhibits
power law fall off at  wide angles precisely due to hard (UV) processes 
\be
A_{qcd}(s,t) \sim (\frac{1}{\sqrt{\alpha'_{qcd} s}})^{n-4}   \; ,
\ee
where $n = \sum_i n_i = \sum_i(d_i - s_i)$ is give as the sum over the twist ($n_i$)
for each external state. In stark contrast the fundamental strings (in flat
space) exhibits exponentially damped wide angle scattering,
\be
A_{closed}(s,t) \rightarrow \exp\big[ - \frac{1}{2} \alpha'(s \ln s + t \ln t
+ u \ln u ) \big] \; .
\ee

Polchinski and Strassler~\cite{Polchinski:2001tt} made the essential
observation on how string scattering in a confining background AdS background
might avoid this conflict with QCD.  Suppose you have a background that is
cut-off for small $r<r_{min}$ and approximated by $AdS^5\times X_5$ for large
r,
\be
ds^2 = \frac{r^2}{R^2} \eta_{\mu\nu} dx^\mu dx^\nu + \frac{R^2}{r^2}
dr^2 + ds^2_X\;.
\ee
A plane wave external glueball ($\phi(r) \exp[ixp]$) at strong coupling scatters locally 
in r through a string amplitude with a red shifted proper distance
or equivalently an effective momenta,
$$
\hat p_s(r) = \frac{R}{r} p \; .
$$
Relative to the string scale, $l_s = \sqrt{\alpha'}$, the
exponential cut-off at high  momenta ($l_s p_s > 1$), suppresses
string scattering in the IR region ($r < r_{scatt}$), leaving a residual
amplitude in a decreasingly small window in the UV ($l_s p_s < 1$),
$$
r > r_{scat} \equiv \sqrt{\alpha'} R p \; .
$$
Since the tail of the glueball wavefunction, $\phi_i(r) \sim
(r/r_{min})^{-\Delta^{(i)}_4} $, is entirely determined in 
the String/Gauge dictionary by the conformal weight
$\Delta^{(i)}_4$ of the corresponding gauge operator dual to the string
state, one is led back to the standard parton or naive
dimensional analysis result used  in the wide angle power
counting,
\be
\phi_i(r_{scat})  \sim 
\Big (\frac{r_{scat}}{r_{min}} \Big )^{-\Delta^{(i)}_4} 
\sim    ( \sqrt{\alpha'_{qcd}}\; p )^{- \Delta^{(i)}_4} \; ,
\label{eq:ads5}
\ee
where we have converted to the hadronic scale, $\alpha'_{qcd} \sim
(R/r_{min})^2 \alpha'$ .

In the corresponding M-theory construction (sometime referred to as M-QCD),
all of this appears to be upset because the scaling of the wave function in
$AdS^7$ changes.  For example the scalar glueball with interpolating field
$Tr[F^{\mu\nu} F_{\mu\nu}]$ in $AdS^5$ has $\Delta_4 = 4$ as expected but in
$AdS^7$ the wavefunction scales with $\Delta_6 = 6$ at large r. As pointed out
by Brower and Tan~\cite{Brower:2002er}, this apparent conflict with partonic
expectations is avoided when one realizes that from an M-theory perspective,
strings are a consequence of membranes wrapping the ``11th'' dimension and
that in $AdS^7$ this 11th dimension is warped just like another spatial
coordinate ($x^\mu$) with the proper size: $\hat R_{11}(r) = (r/R) R_{11}$.
To account for this effect, one can introduce local values for the
effective string length and coupling constant,
$$\hat l^2_s(r) = \frac{R}{r} (l^3_p / R_{11})\;, \quad {and} \quad
\hat g^2_s(r) = \frac{r^3}{R^3} (R^3_{11}/l^3_p)\;.$$
as a function of the local scattering position in r.  This additional
deformation is precisely what is required.  The new definition of the
scattering region at wide angles,
$$
 r > r_{scat} = \hat l_s(r_{scat}) R\; p =
\sqrt{\alpha'} \; R^\frac{2}{3} \; r_{scat}^{-\frac{1}{2}} \; p\; , 
$$
leads to
\be
\phi_i(r_{scat}) \sim \Big (\frac{r_{scat}}{r_{min}} \Big )^{-\Delta^{(i)}_6} 
\sim  \Big( \sqrt{\alpha'_{qcd}}\; p \Big)^{-\frac{2}{3} \Delta^{(i)}_6}
\label{eq:ads7}
\ee for each external line.  For example, for the $0^{++}$ scalar glueball
corresponding to interpolating YM operator $Tr[F^2]$, the factor of $2/3$
exactly compensates for the the shift in the conformal dimension from
$\Delta_4 = 4$ for $AdS^5$ to $\Delta_6 = 6$ for $AdS^7$ to give the parton
results, $n_i = \frac{2}{3}\Delta^{(i)}_6$.  This time, in converting to the
hadronic scale in Eq.~\ref{eq:ads7}, we must realize the relationship of
$\alpha'_{qcd}$ to the string scale is \be \alpha'_{qcd} \sim (R/r_{min})^3
\alpha' \; , \ee which differs from the $AdS^5$ string relation
(\ref{eq:alpha5}).  The 3rd power is a consequence of the fact that in
M-theory the area law for the Wilson loop really comes from a minimal volume
for a wrapped membrane world volume stabilized at $r \simeq r_{min}$ rather
than a minimal world surface area for a string which gave quadratic behavior
in Eq.~\ref{eq:alpha5} .

Putting all factors together, the result for M-theory  can be expressed 
as,
$$
\Delta \sigma_{2 \rightarrow m} \simeq \frac{1}{s}\;f(\frac{p_i \cdot p_j}{s}) 
 \frac{1}{N^{2m}} \;\prod_i \Big( \frac{1}{\alpha'_{qcd} s} \Big)^{n_i
 - 1} \;,
$$
in correspondence with the weak-coupling QCD result.

Summarizing the results on hard scattering:
\begin{enumerate}
\item $AdS^5$ by Polchinski-Strassler:\hfil
$
\Delta \sigma_{2 \rightarrow m} \simeq \frac{1}{s}\;f(\frac{p_i \cdot p_j}{s}) 
 \frac{(\sqrt{g^2 N_c})^m}{N_c^{2m}} 
\;\prod_i \Big( \frac{1}{\alpha'_{qcd} s} \Big)^{n_i - 1} 
$
\item $AdS^7$ by Brower-Tan:\hfil
$
\Delta \sigma_{2 \rightarrow m} \simeq \frac{1}{s}\;f(\frac{p_i \cdot p_j}{s}) 
 \frac{1}{N_c^{2m}} \;\prod_i \Big( \frac{1}{\alpha'_{qcd} s} \Big)^{n_i - 1} 
$
\item QCD perturbation theory:\hfil
$
\Delta \sigma_{2 \rightarrow m} \simeq \frac{1}{s}\;f(\frac{p_i \cdot p_j}{s}) 
 \frac{(g^2N_c)^m}{N_c^{2m}} \;\prod_i \Big( \frac{g^2N \Lambda_{qcd}^2}{\alpha'_{qcd} s} \Big)^{n_i - 1} 
$
\end{enumerate}

\subsection{Near-Forward Scattering and Regge Behavior}

The importance of scattering at large r also implies the presence of a
hard component in the near-Regge limit, $t/s \rightarrow 0$ as
$s\rightarrow \infty$.  The approximation of a single local scattering
leads to $ T(s,t) = \int_{r_h}^{\infty} dr \;{\cal K}(r) A(s, t,
r), $ where $A$ is a local 4-point amplitude, ${\cal K}(r) \sim r^5
\phi_1(r) \phi_2(r) \phi_3(r) \phi_4(r)$, up to a constant, and $r_h$
is a cut-off, $r_h>>r_{min}$.  After converting to local string
parameters as discussed above, the amplitude $A(s, t, r)$ depends only
on $\alpha' \hat s$ and $\alpha' \hat t$, where $\hat s
= (R/r)^3 s$ and $\hat t = (R/r)^3 t$. In the Regge limit the
amplitude becomes

\be 
T(s,t) = \int_{r_{h}}^{\infty} dr\; {\cal K}(r)  \beta( \hat t)
(\alpha' \hat s)^{\alpha_0 + {\alpha'}  \hat t }\; .
\label{eq:regge}
\ee
For small $t\simeq 0$, this corresponds to an exchange of a BFKL-like Pomeron, 
with a  small effective
Regge slope,
\be \alpha'_{BFKL}(0)\sim
 (r_{min}/r_h)^3\alpha'_{qcd}<< \alpha'_{qcd} \; .  
\ee
Such an exchange naturally leads to an elastic diffraction peak with
little shrinkage. In the coordinate space, one finds, for a hard
process, the transverse size is given by
\be 
\<{\vec X}^2\> \sim (r_{min}/r_h)^3 \alpha'_{qcd} \log s +{\rm constant} \; . 
\label{eq:xsq} 
\ee 
If the cutoff, $r_h$, which characterizes a hard process, increases
mildly with $s$, e.g. $r_h^3\sim \log s$, there will be no transverse
spread.  In the language of a recent study by Polchinski and
Susskind,~\cite{Polchinski:2001ju}, this corresponds to ``thin" string
fluctuation.

In spite of this progress in seeing some partonic effects in the
string picture, there is much more to understand. For instance, we
note that, consistent with the known spectrum of glueballs at strong
coupling, the IR-region must in addition give a factorizable Regge
pole contribution,
\be 
T(s,t)\sim A( s, t, r_{min}) \sim (\alpha'_{qcd} s ) ^ {\alpha_{P}(0) + \alpha'_{qcd} t
} \; .
\ee
In our M-theory construct~\cite{Brower:2000rp}, $\alpha_P(0)=2-0(1/g^2N)$. Of
course, this ``soft" Pomeron must mix with the corresponding hard component,
leading to a single Pomeron singularity in the large N limit.  However,
addressing this issue requires a more refined treatment for the partonic
structure within a hadron. As emphasized by Polchinski and Strassler in a
recent paper~\cite{Polchinski:2002jw}, this is also what is required for
treating deep inelastic scattering in the string/gauge duality picture.
Efforts in this direction are currently underway.

\section{Lecture Three: String vs Lattice Spectra }

Based on the conformally broken backgrounds using Maldacena string/gauge
duality, we can begin to do some calculation in QCD like theories, at least
in the strong coupling limit.  This is still far from the hoped for discovery
of {\bf the} QCD string. We are in the position somewhat similar to a lattice
cut-off theory. The strong coupling limit brings along non-universal cut-off
dependent effects. However unlike the lattice, we have (as yet) no algorithm
(theoretical or numerical) to in principle send the cut-off to infinity.
String/gauge duality presents a coupled problem, even in the large $N_c$.
The world sheet sigma model for the string theory  emits gravitons that
perturb the background which in turn has a back reaction on the sigma model.
Even finding the sigma model beta function perturbatively to the next order in
$1/\alpha'$ is difficult.  Still it is worth while to see if there is a
reasonable spectrum in strong coupling approximation.

On the lattice side, where one can numerically take the weak coupling
(continuum) limit, the spectra for glueballs and the quantum states of a
stretch string are becoming quiet accurately determined. Even extrapolating
this spectra to the large $N_c$ limit has met with some success. In short the
lattice has given and is capable of giving more accurate spectral data for
the quantum QCD string. If it exists, there can be only one
answer.  This is a unique opportunity: A concrete string theory problem with
copious ``experimental'' data to constrain its construction!

\subsection{Glueball Spectra}

\begin{figure}[ht]
\raisebox{0.42cm}{\includegraphics[width=2.75in,height=4.0in]{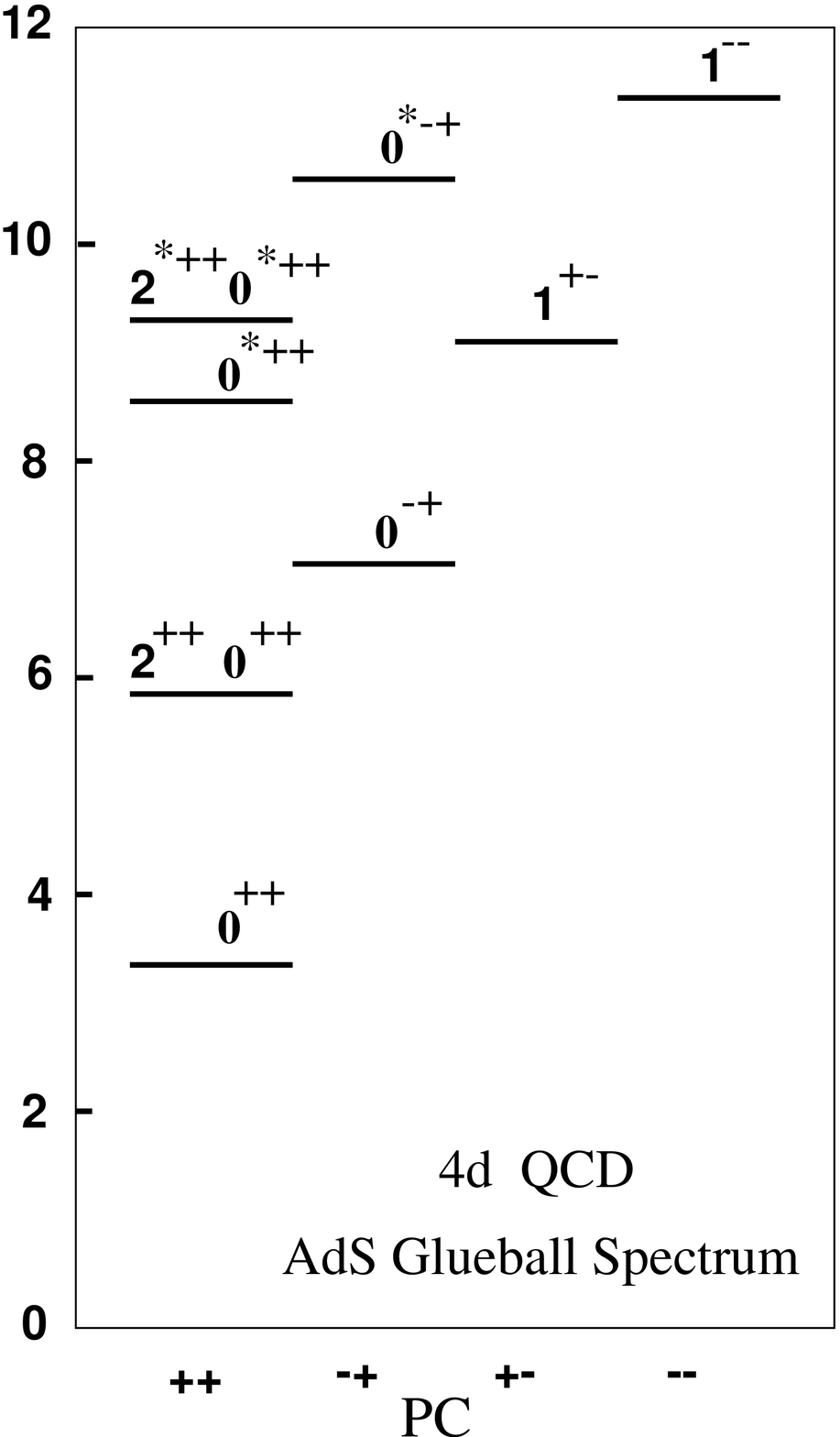}}
\includegraphics[width=3.25in,height=4.20in]{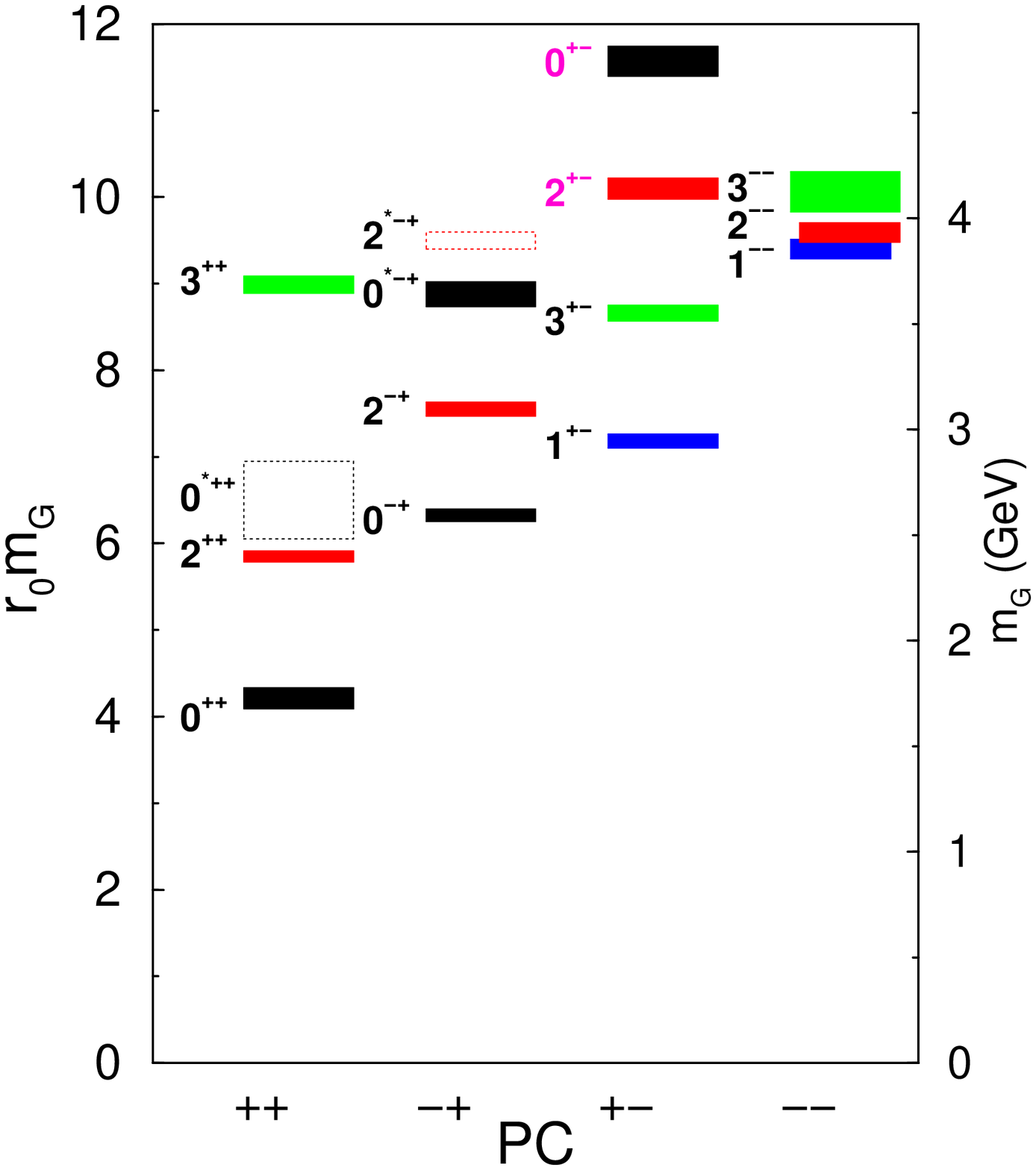}
 \caption{The AdS glueball spectrum~\cite{Brower:2000rp} for $QCD_4$ in strong
 coupling (left) compared with the lattice spectrum~\cite{MP} for pure
 SU(3) QCD (right). The AdS cut-off scale is adjusted to set the
 lowest $2^{++}$ tensor state to the lattice results in units of
 the hadronic scale $1/r_0 = 410$ Mev.}  
\label{fig:comparison}
\end{figure}

The first such lattice AdS/CFT comparison was the computation of the  strong coupling 
glueball spectrum in the $AdS^7$ M-theory black hole. 
The correspondence for the  quantum numbers for the gravity modes
in terms of the Yang-Mill fields are read off the effective Born-Infeld
action on the brane,
\be
S=\int d^5x
\det[G_{\mu\nu}+e^{-\phi/2}(B_{\mu\nu}+F_{\mu\nu})]+\int d^4x (C_1
F\wedge F+ C_3\wedge F+ C_5)
\ee
The entire spectrum for all states in the QCD super selection
sector are now known and can be compared with lattice data for
SU(3). The comparison is rather encouraging considered as a first
approximation (see Fig.~\ref{fig:comparison}). All the states are in
the correct relative order and the missing states at higher J
are a direct consequence of strong coupling which pushes the string
tension to infinity.
It appears plausible that the $AdS^7$ black hole phase at strong
coupling is rather smoothly connected to the weak coupling (confined)
fixed point of QCD.  However it must be stated that there is no
general understanding of how the metric will be deformed so that all
the unwanted charged Kaluza-Klein states in the extra compact
directions decouple. All attempts to find better background solutions
to supergravity as a starting point for QCD have failed in this
regard.

\subsection{Stretched String Spectra}

\begin{figure}[ht]
\begin{center}
\includegraphics[width=4.0in,height=4.0in]{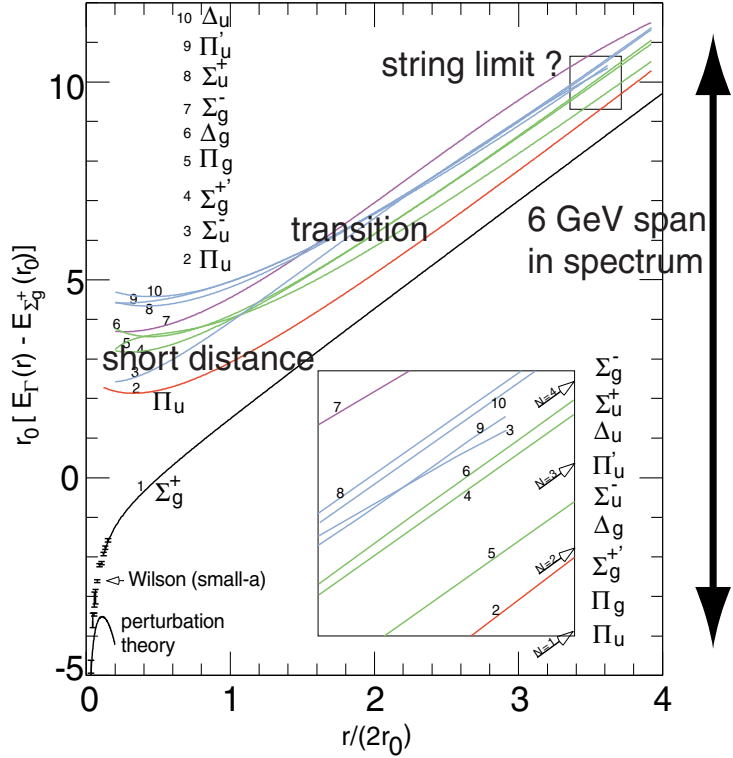}
\caption{Quantum fluctuation of a stretch QCD flux tube~\cite{kuti}}
\label{fig:stringspec}
\end{center}
\end{figure}

An even more direct observation of the string spectrum in lattice gauge data
is giving by the quantum modes of a stretch string between fixed infinitely
heavy sources (see Fig~\ref{fig:stringspec}). This is the open QCD string
with Dirichlet boundary conditions. From the $AdS/CFT$ view point starting
with the string ends separated by a small distance L, we are able to see
first the short distance coulomb regime. Then as we increase L, the minimal
surface moves into the interior probing more and more IR physics. Finally at
very large L, we see only the $O(1/L)$ low mass   pseudo Goldstone modes 
for the transverse co-ordinates of the string leading to the universal
spectrum of L\"uscher,
\be
E =  T_0 L
 - \frac{\pi(D-2)}{12  L} + \frac{2 \pi a^\dagger_n a_n}{L} + \cdots
\ee
Indeed at large separation L the lattice data for the stretched string
spectrum appears to be approaching this form with $D - 2 = 2$ transverse
oscillators (See fig~\ref{fig:stringspec}). 

Very careful and clever methods have been develop by L\"uscher and
Weisz~\cite{Luscher:2002qv} to determine the universal ``L\"uscher term''
giving a fit in the range 0.5 to 1.0fm to the ground state (i.e. static
potential):
$$E_0(L)  = \frac{\pi}{12}(1 + 0.12fm/L)$$
confirming this prediction.

A major challenge to the AdS/CFT approach to the QCD string is to understand
the interpolation between large L and small L. If we take as a model, the 
``warped'' metric,
\be
ds^2 = V(y) dx^\mu dx_\mu + dy^2 + U(y) d^2\tau + \cdots
\ee
suggested by an $AdS^{d+2}$ black hole with $ V(y) = r^2/r^2_{min} =
[cosh(\frac{d+1}{2} k y)]^{4/(d+1)} $. We can solve (numerically) for the
classical minimal surface getting a potential energy, $E_0(L)$, for the 
classical ground state stretched string. For $d=3$ and
$R^4 = 2g^2_{YM} N_c \alpha'^2$, it  has the limiting values,
\be
E_0 \rightarrow \frac{r^2_{min}}{2 \pi R^2 \alpha'} \; L    + O(L  e^{- c L}) \tbox{and}
E_0 \rightarrow - \frac{(2\pi)^2}{\Gamma(1/4)^4 } \;  
                  \frac{\sqrt{2 g^2_{YM} N_c}}{L} 
\ee 
for $L\rightarrow \infty$ and $L\rightarrow 0$ respectively.  The shape of
the potential,$E_0(L)$, fits the lattice data almost perfectly after setting
the QCD length scale ($r_{min} = /\Lambda_{qcd}$) and string tension
($T_{qcd} = 1/2 \pi \alpha'_{qcd} = r^2_{min}/2 \pi R^2 \alpha'$).  This is
reassuring but also highlights the present situation. QCD itself in the
continuum limit will give a definite number for the string tension, $T_{qcd}$
(at large N) relative to the QCD scale, $\Lambda_{qcd}$.  But at strong
coupling in the AdS/CFT (or on the strong coupling lattice for that matter),
there is an extra parameter that allows these to fixed independently.  One must
flow to the asymptotically free UV fixed point to eliminate this parameter.

Work is currently underway to study the fluctuations as well in this model
background~\cite{Brower:2005pb}. In a gauge with $\sigma = z = X^3$ the
transverse fluctuation obey the equation,
\be
[\; \partial^2_t - v^2(z) \partial^2_\sigma\; ]\; X_\perp = 0 \; ,
\ee
and the radial mode,
\be
[\; \partial^2_t - v^2(z) \partial^2_\sigma\; ]\; \xi = M^2(z) \xi \; ,
\ee
with $v^2(z) = V^2(0)/V^2(z)$ and $M^2(z) =V^2(0)[\; V''(z)/V^2(z) - 3
V'^2(z)/2 V^3(z)\;] \simeq const M^2_{BG}$ choosing the string to stretch
symmetrically in the interval $z \in [-L/2, L/2]$. The local velocity of
waves on the string $v(z)$ is bounded by the speed of light, slowing as it
approaches the infinitely massive quarks at the ends. It is clear that at
large L this will reproduce the L\"uscher result for a D-2 = 2 string. In
addition there are quantum modes for fluctuations in the extra ``radial''
directions with a rest mass set by the glueballs. These modes correspond
through string/gauge duality to longitudinal modes for a fat chromodynamic
flux tube. However this toy QCD string is at best just a qualitative model of
how a QCD string in warped space might behave. To find the microscopic
degrees of freedom on the world sheet for the QCD string will require
a detailed understanding of how high frequencies are governed by the short
distance properties of asymptotically free gauge theory at large $N_c$.

\section{No Conclusions Yet}

The construction of the QCD string theory remains a tantalizing but
unrealized goal. Recent progress has certainly begun to show how such an
exact string/gauge duality may arise. Indeed the intimate relations between
Yang-Mills theory and string theory is a dramatic change in our
understanding, which may aptly designated the ``First String Counter
Revolution'' -- bring the subject back to its earliest roots. In these short
lecture notes it has not been possible to describe fascinating new insight
into issues concerning the introduction of dynamical quarks, spontaneous
chiral symmetry breaking, non-perturbative effects in the coupling $g_s \sim
1/N_c$ such as the giant graviton baryon connection and as well as attempts
to reach short distance physics from the string description. However there is
still much confusion on each of these topics with new ideas streaming forth.
The most definitive mathematical progress based on String/Gauge duality is in
tractable ``toy models'' of QCD with some residual Supersymmetry or special
limits where semi-classical methods can be applied.

It must also be admitted that formidable challenges remain. Even for the
simplest case of pure $AdS^5 \times S^5$, it has not been possible to
analytically quantize the free superstring. Hard evidence for AdS/CFT duality
is often somewhat indirect requiring examples with strong constraints from
supersymmetry or unphysical limits with high Kaluza-Klein charges.  A basic
problem remains to find a mechanism to really separate the charged
Kaluza-Klein state outside the QCD sector from the physical states that
should survive at a QCD fixed point.  Perhaps the top down framework of
starting from a critical supersymmetric string is flawed.  Bottom up methods
starting with non-critical strings with no supersymmetry and few or no
compact dimensions are difficult but worth pursuing. However at present there
is no a direct constructive method for defining {\bf the} QCD string dual to
pure Yang Mills theory even at $N_c =\infty$, ignoring the subsequent
difficulty in solving it.  This is in contrast with lattice gauge theory
which is well defined in spite of the need at present to resort to numerical
methods for its solution.  Nonetheless the ancient conjecture that QCD is in
fact dual to a fundamental string theory is more plausible and we are
finding more and more about how such dualities arise.  Let's hope that a
young ``Veneziano'' comes to the rescue early in this Millennium.

{\bf Acknowledgments:}

I would like to acknowledge the fine hospitality of my friends at the Cracow
School of Theoretical Physics who gave me the opportunity to try out these
lectures at Zakopane in 2003 and my colleague Chung-I Tan who contributed
in enumerable ways to this presentation.


\newpage
\begin{thebibliography}{9}

  
\bibitem{note}WARNING: This is not intended as a history of early string
  theory or a review of modern developments.  A fair or even moderately
  complete treatment of either does not lend itself to three short
  pedagogical lectures.



\bibitem{Maldacena:1997re}
J.~M.~Maldacena,
``The large N limit of superconformal field theories and supergravity,''
Adv.\ Theor.\ Math.\ Phys.\  {\bf 2}, 231 (1998)
[Int.\ J.\ Theor.\ Phys.\  {\bf 38}, 1113 (1999)]
[arXiv:hep-th/9711200].

\bibitem{Aharony:1999ti} For a review of the classsic AdS/CFT framework see
O.~Aharony, S.~S.~Gubser, J.~M.~Maldacena, H.~Ooguri and Y.~Oz,
``Large N field theories, string theory and gravity,''
Phys.\ Rept.\  {\bf 323}, 183 (2000)
[arXiv:hep-th/9905111].


\bibitem{BrowerHarte} R. C. Brower and J. Harte, ``Kinematic Constraints
for Infinitely Rising Regge Trajectories'', Phys. Rev. {bf 164} (1967) 1841.


\bibitem{Dolen:1967jr}
  R.~Dolen, D.~Horn and C.~Schmid,
  ``Finite Energy Sum Rules And Their Application To Pi N Charge Exchange,''
  Phys.\ Rev.\  {\bf 166}, 1768 (1968).


\bibitem{veneziano} G. Veneziano, ``Construction of a crossing-symmetric
  Regge-behaved amplitude for linearly rising trajectories'', Nuovo Cim 57A
  (1968) 190-197. To be precise this paper gave an amplitude for $\omega
  \rightarrow \pi^+ \pi^- \pi^0$ which was extended to pion scattering  by
  Lovelace and Shapiro shortly thereafter.


\bibitem{Neveu:1971rx}
  A.~Neveu and J.~H.~Schwarz,
  ``Factorizable Dual Model Of Pions,''
  Nucl.\ Phys.\ B {\bf 31}, 86 (1971).

  
\bibitem{Polchinski:1998rq}, J. Polchinski, "String theory. Vol. 1: An
  introduction to the bosonic string", Cambridge, UK: Univ. Pr. (1998)
Superstring theory and beyond", Cambridge, UK University. Press. (1998).

\bibitem{Zwiebach:2004tj}
For more details on classical solution and an excellent introduction to
string theory see,
B. Zwiebach ``A first course in string theory'',
Cambridge, UK: Univ. Pr. (2004)


\bibitem{Polchinski:1998rq2} A proper generalization
of the dual pion amplitude is the superstring, which is
technically more involved.  see J. Polchinski, "String theory. Vol. 2:

\bibitem{thooft} G. 'tHooft, ``A planar diagram theory for strong interactions'',
Nuicl. Phys. B 72 (1974) 461-473.


\bibitem{Armoni:2003fb}
  A.~Armoni, M.~Shifman and G.~Veneziano,
  ``SUSY relics in one-flavor QCD from a new 1/N expansion,''
  Phys.\ Rev.\ Lett.\  {\bf 91}, 191601 (2003)
  [arXiv:hep-th/0307097].



\bibitem{Polchinski:2001tt}
J. Polchinski and  M. Strassler ``Hard scattering and gauge/string duality'',
Phys. Rev. Lett. 88, 031601 (2002)

\bibitem{Brower:2002er}
R. C. Brower and C-I Tan ``Hard scattering in the M-theory dual for the QCD string''
Nucl. Phys. B662,393-405 (2003)


\bibitem{Polchinski:2001ju}
  J.~Polchinski and L.~Susskind,
  ``String theory and the size of hadrons,''
  arXiv:hep-th/0112204.

\bibitem{Brower:2000rp}
  R.~C.~Brower, S.~D.~Mathur and C.~I.~Tan,
  ``Glueball spectrum for QCD from AdS supergravity duality,''
  Nucl.\ Phys.\ B {\bf 587}, 249 (2000)
  [arXiv:hep-th/0003115].


\bibitem{Polchinski:2002jw}
  J.~Polchinski and M.~J.~Strassler,
  ``Deep inelastic scattering and gauge/string duality,''
  JHEP {\bf 0305}, 012 (2003)
  [arXiv:hep-th/0209211].


\bibitem{MP} C. J. Morningstar and M. Peardon, ``The glueball spectrum from an
anisotropic lattice study'', Phys.Rev. D60 (1999) 034509.

\bibitem{kuti} K. Jimmy Juge, Julius Kuti and Colin Morningstar
``The QCD String Spectrum and Conformal Field Theory''
Nucl.Phys.Proc.Suppl. 106 (2002) 691-693.


\bibitem{Luscher:2002qv}
  M.~Luscher and P.~Weisz,
  ``Quark confinement and the bosonic string,''
  JHEP {\bf 0207}, 049 (2002)
  [arXiv:hep-lat/0207003].


\bibitem{Brower:2005pb}
R.~C.~Brower, C.~I.~Tan and E.~Thompson,
``Probing the 5th dimension with the QCD string,''
arXiv:hep-th/0503223.
  
\end{thebibliography}
\end{document}